\documentclass[preprint]{elsarticle}
\usepackage{bm}
\usepackage[T2A]{fontenc}
\usepackage[cp1251]{inputenc}
\usepackage[english]{babel}
\usepackage{amsfonts,amssymb,amsmath}
\usepackage{amsmath,graphicx}
\usepackage{graphicx,pifont} 
\usepackage{epsfig}
\title{On physical principles and mathematical mechanisms of the phenomenon of irreversibility}
\author[ayz]{A.Yu.~Zakharov$ ^{*} $} 
\ead{Anatoly.Zakharov@novsu.ru}

\cortext[ayz]{Corresponding author}

\fntext[fn1]{The work was carried out with financial support of the Ministry of Education and Science of Russian Federation within the framework of the project part of the state order (project No.~3.3572.2017).}
\address{Yaroslav-the-Wise Novgorod State University, Veliky Novgorod, 173003, Russia}
 \begin{document}
\begin{abstract}
It is shown that the phenomenon of irreversibility in many-body and few-body systems can be explained and described within the framework of the concept of direct (not instantaneous) interaction of particles without using probabilistic hypotheses. The exact solution of the model of a two-particle classical oscillator with retarded interaction between particles is presented. It is established that the interactions retardation leads to appearance of an infinite spectrum of both stationary and non-stationary oscillations, and to  non-invariance of the solution with respect to time reversal as well.

\end{abstract}


\maketitle
\section{Introduction}
At present, there is no real alternative to statistical mechanics, which is the basis of theoretical study of many-particle systems, including the theory of non-ideal gases, plasma, condensed matter physics, etc. It is considered that the problem of calculating of the thermodynamic properties of a substance is reduced to purely mathematical problems such as finding of partition functions or uncoupling Bogolyubov-Born-Green-Kirkwood-Yvon hierarchies. However, there are very significant actual problems of statistical mechanics, which are related not to the computational part, but to the problem of its substantiation. The most significant of these problems are:
\begin{enumerate}
	\item 
	First of all, this is the problem of consistency of the combination of deterministic reversible classical mechanics with probabilistic assumptions within the framework of classical statistical mechanics. In particular, the paradoxes (in fact, the contradictions) of Loshmidt and Zermelo testify to the internal contradictions of such an integration. The nature of the Zermelo paradox was clarified in the work of Kac~\cite {Kac}, in which a dynamic model of a many-particle system (the ring model) was proposed and two solutions of this model were obtained. The first solution is an exact dynamic solution that is reversible in time and, moreover, periodically, as one would expect in connection with the Poincar\'e recurrence theorem. The second solution of the ring model was obtained by Kac using highly plausible probabilistic hypotheses. This solution has the property of irreversibility in time. These two solutions are almost identical at small time intervals, but differ significantly on large time scales. Thus, the use of probabilistic concepts without establishing of a real physical mechanism cannot be recognized as a microscopic justification of thermodynamics and an explanation of the observed thermodynamic behavior of real systems. Moreover, this means that the explanation or justification of thermodynamics is outside the score of classical mechanics. Therefore, the use of exact results of classical mechanics, including the Liouville theorem on the conservation of phase volume and the Liouville equation for the evolution of distribution functions, does not seem convincing.
	\item 
	Even assuming  the legitimacy of the probabilistic approach to explain the thermodynamic behavior, it is necessary to exclude from consideration the exact results of classical mechanics (such as the Liouville, Poincare theorems, etc.), and instead to introduce the methods for calculating both the probabilities of the system states and the transition probabilities between these states . With an unknown mechanism for stochastization of the system, it is impossible to justify the method for calculating of these probabilities. Therefore, the values of these probabilities cannot be calculated, and they should only be postulated. In particular, the principle of equal probabilities in the Gibbs microcanonical distribution, methods for calculating of transition probabilities in the Boltzmann kinetic equation, etc. are postulated. In this regard, we note that the applicability of the limit theorems of the probability theory used in statistical mechanics is restricted by the condition of independence of systems of random variables~\cite{Prokhorov,Petrov}. However, the interaction between the particles means the unconditional exclusion of their independence. In other words, the rigorous mathematical foundation of statistical mechanics by means of probability-theoretic assumptions~\cite{Khinchin1,Khinchin2} is applicable at best to systems of noninteracting particles, i.e. to the ideal gases only. 
	
	In addition, it should be noted that the limits of equivalence for Gibbs statistical ensembles (microcanonical, canonical, and grand canonical distributions) are not completely defined. Therefore, the results of solving of specific problems may be significantly dependent on the type of ensemble used in the solution. For example, metastable states do not exist within a grand canonical distribution, while they can be quite studied within a canonical distribution. 
	
	Finally, in the canonical distribution, phase transitions exist only for infinite systems (that is, only under the condition that the thermodynamic limit transition is performed)~\cite{Uhlenbeck}. On the contrary, numerical studies in the framework of the microcanonical ensemble reveal the signs of phase transitions in finite systems~\cite{Gross}. From this point of view, statistical mechanics as a whole is not quite well-posed theory.
\end{enumerate}

\section{On physical principles of the phenomenon of irreversibility}

In 1909, a discussion paper by Ritz and Einstein~\cite{Ritz} was published, in which the authors presented the directly opposite views on the problem of radiation: ``Ritz considers the restriction in the form of retarded potentials as one of the sources of the second law of thermodynamics, whereas Einstein assumes that irreversibility is based exclusively on the probabilistic foundations''. Currently, the probabilistic point of view~\cite{Ehrenfest, Tolman,Landau} is dominant in the theory of many-body systems, and the study of these systems is based on the field theory of interactions between particles.
In this approach, the evolution of classical systems is described by the equations of motion of both the particles and the fields, and the interaction between the particles is carried out through the field. 

At first sight, it may seem that within the framework of the field theory of the picture of interactions between particles, the problem of the irreversibility origin remains open, since the equations of motion of both particles and fields are invariant with respect to the time reversal $ t \to -t $. However, this is not true.

The fact is that the solutions of the field equations of motion consist of two contributions. The first contribution is a general solution of homogeneous field equations that describe the free fields. These are the fields that exist in the absence of field sources, i.e. charges and currents. 
The second contribution is advanced and retarded potentials, which are solutions of inhomogeneous field equations with given charges and currents. However, advanced potentials violate the principle of causality and therefore should be excluded from consideration. In other words, it is necessary to select only those solutions of field equations that correspond to retarded potentials of such type, as the Li\'{e}nard–Wiechert potentials in electrodynamics.
Thus, the principle of causality and the existence of an upper limit of speed — the speed of light — lead to appearance of a particular direction of time, despite the symmetry of the field equations with respect to the time reversal.

It is currently not known whether it is possible to transform field equations of general type to a form that is not invariant with respect to the time reversal.
Therefore, it is of interest to study the dynamics of systems of interacting particles in the framework of an approach that does not use the field equations. One of the variants of this approach is the theory of direct interaction between particles.

The theory of direct interaction between particles as an alternative to the field theory of interactions is based on the works of Wheeler and Feynman~\cite {Wheeler1, Wheeler2}, who established the equivalence of the field theory of electromagnetic interaction between particles and the theory of direct (not instantaneous) interaction.

Although the interaction between molecules is of an electromagnetic nature, the ``real'' intermolecular potentials may differ significantly from Coulomb's law. We denote the interaction potential of two \textit{resting} particles, located at the points $ \mathbf{r} $ and $ \mathbf{R} $ by a function $ W \left( \mathbf{r} - \mathbf{R} \right) $. Then the total potential created at the point $ \mathbf{r} $ by a system of particles distributed in space with the microscopic density
\begin{equation}\label{n(r,t)}
n\left(\mathbf{r},t \right) = \sum_{s=1}^{N}\delta\left(\mathbf{r}-\mathbf{R}_{s}\left(t \right)  \right), 	
\end{equation}
has the following form
\begin{equation}\label{Lienard1}
\varphi\left( \mathbf{r}, t\right) = \int  W\left(\mathbf{r} - \mathbf{r}' \right) n\left(\mathbf{r}', t - \tau\left(\left| \mathbf{r} - \mathbf{r}'\right|  \right)  \right)\  d\mathbf{r}',
\end{equation}
where $ \tau\left(\left| \mathbf{r} - \mathbf{r}'\right|  \right)  $~is the retardation time of interactions between points $ \mathbf{r} $ and $ \mathbf{r}' $:
\begin{equation}\label{tau1}
\tau\left(\left| \mathbf{r} - \mathbf{r}'\right|  \right) = \frac{\left|\mathbf{r} - \mathbf{r}'\right| }{c},
\end{equation}
$ c $~is the speed of propagation of interactions (the speed of light).

The dynamics of systems in the framework of the theory of direct retarded interaction of particles is described by functional differential equations, a special case of which are differential-difference equations.

The delay of the interactions leads to violation of Newton's third law on the equality of action and reaction~\cite{Ivlev}, and also it is one of the mechanisms for the irreversible behavior of many-body systems~\cite{Zakh2,Zakh3}. 
Note that in papers~\cite{Zakh2,Zakh3} neither probabilistic considerations nor the conditions $N \gg 1$ ($N$~is number of degrees of freedom in the system) were used. In this connection, it is interesting to study in detail the mechanisms of manifestation of the interactions retardation  in irreversibility by the example of a system with a small number of degrees of freedom.

As a suitable few-body system, let us consider a system of two particles with the same masses $ m $, the interaction between which \textit{at rest} is described by the potential $ W \left(\mathbf{R}_1 - \mathbf{R}_2 \right) $ with the following properties.
\begin{enumerate}
	\item The function $W\left(\mathbf{R}_1 - \mathbf{R}_2 \right)$ has a minimum at  $\left|\mathbf{R}_1 - \mathbf{R}_2 \right| = L$.
	\item Near the minimum point, this function can be approximated by the quadratic function
	\begin{equation}\label{kx2}
	W\left(\mathbf{R}_1 - \mathbf{R}_2 \right) = W\left(L \right) + \frac{k\left(\left|  \mathbf{R}_1 - \mathbf{R}_2\right| - L \right) ^2}{2}.
	\end{equation}
\end{enumerate}

Let us consider the case of one-dimensional \textbf{small oscillations} of these particles along a straight line connecting these bodies.

In the absence of retardation, the solution of this problem is trivial and describes the oscillations of the particles with circular frequency $\displaystyle \sqrt{\frac{2k}{m}} $.
The frequency of these oscillations does not depend on either $L$ or $W\left(L \right) $. The retardation in the interactions between these two particles leads to the qualitatively different results, including the irreversible behavior of the system as a whole.

\section{Oscillator with retardation}

We denote the deviations from the equilibrium positions of the particles in the rest state by $x_1\left( t\right) $ and $x_2\left( t\right) $. 
Equations of motion for the system in view of the interaction retardation~$\tau$  have the form: 
\begin{equation}\label{Eq-2}
\left\lbrace 
\begin{array}{l}
{\displaystyle \ddot{x}_1(t)\, + \omega_0^2 \left[x_1(t) - x_2(t-\tau) \right] = 0; } \\
{\displaystyle \ddot{x}_2(t)\, + \omega_0^2 \left[x_2(t) - x_1(t-\tau) \right] = 0,}
\end{array}
\right.
\end{equation}
where $\omega_0 = \sqrt{\frac{k}{m}}$.

In general, $ \tau $, of course, is not a constant, but a function of the position of the particles, depending on the solution of the system of equations~\eqref{Eq-2}. The analytical solution of such a problem in general formulation is far beyond the capabilities of modern mathematics.

The condition of smallness of the oscillations
\begin{equation}\label{small}
\left| x_k\left(t \right) \right| \ll L, \quad (k=1, 2) 
\end{equation} 
simplifies the problem: in this case the principal term of the interaction retardation is reduced to the constant
\begin{equation}\label{tau}
\tau = \frac{L}{c},
\end{equation}
Thus, in the case of small oscillations, the system of equations~(\ref{Eq-2}) is linear.

The Euler substitution 
\begin{equation}\label{Euler}
x_{k} \left(t\right)=C_{k}\, e^{i\, \omega_0\, \Lambda t}
\end{equation}
leads to the following characteristic equation with respect to dimensionless frequency $ \Lambda $:
\begin{equation}\label{char2}
\Lambda ^{4} -2\Lambda ^{2} + \left[1-e^{-2i \omega_0 \Lambda \tau } \right]=0. 
\end{equation}
The roots of this equation depend on the magnitude of interactions retardation~$ \tau $, and finally on the parameter $ L $. Generally, the equation~\eqref{char2} is transcendental, therefore the number of its roots is infinite and the roots are complex. Thus, it would seem that the quantitatively insignificant effect of the interactions retardation belongs to the singular perturbations, which leads to a radical rearrangement of the solutions of the equations. Note that in the absence of a retardation $ \tau = 0 $, the equation~\eqref{char2} is algebraic and therefore has a finite number of roots (4 roots).

We extract real and imaginary parts in $ \Lambda $
\begin{equation}\label{Im-alpha}
\Lambda =  \alpha +i\gamma
\end{equation}
and transform the equation~\eqref{char2} with respect to $ \Lambda $ to a system of equations with respect to $ \alpha $ and $ \gamma $:
\begin{eqnarray}\label{char1}
\left\{\begin{array}{r} 
{\alpha ^{4} -6\alpha ^{2} \gamma ^{2} +\gamma ^{4} -2 \left(\alpha ^{2} -\gamma ^{2} \right)+ 
	\left[1-e^{2\omega_0 \gamma \tau } \cos \left(  2\omega_0\alpha \tau \right) \right]=0;}\\ 
{4\alpha \gamma \left(\alpha ^{2} -\gamma ^{2} - 1 \right)  +  e^{2\omega_0 \gamma \tau } \sin\left(  2\omega_0 \alpha \tau\right) =0.} \end{array}\right.
\end{eqnarray}
Note that $ \alpha $ determines the oscillation frequency, and $ \gamma $ characterizes the rate of change in the amplitude of the oscillations. Therefore, the condition for stationarity of the oscillations is that $ \gamma=0 $, whence we obtain
\begin{equation}\label{b=0}
2\omega_0 \alpha \tau =\pi n,   
\end{equation}
where $n$~is an arbitrary natural number. 

Substituting $ \tau $ from~(\ref{b=0}) and $ \gamma = 0 $ into the first equation of the system~(\ref{char1}), we obtain the equation with respect to~$ \alpha $
\begin{equation}\label{a-n}
\alpha ^{4} -2\alpha ^{2} +2 \sin^2\left( \frac{\pi n}{2} \right) = 0. 
\end{equation}
The roots of this equation are real if and only if $ n $ is an even number. Under this condition, the solutions are as follows:
\begin{equation}\label{a1-a4}
\alpha _{1} =\alpha _{2} =0;\quad \quad \alpha _{3,4} =\pm \sqrt{2},\quad L = \frac{n\lambda}{4},\quad  \lambda=\frac{2\pi c}{\sqrt{2}\omega_0}.
\end{equation}
Thus, stationary oscillations of a two-particle oscillator with retarded interactions occur only for a discrete set of equilibrium distances~$ L $ between the particles, determined by the condition~(\ref {b=0}).

However, there exists a set of values of the parameter~$ L $ for which both stationary and non-stationary oscillations are simultaneously possible, i.e. solution of the system of equations~\eqref{char1} with $\gamma = 0 $ and $\gamma \not = 0 $, respectively. 
From the immense set of solutions of this system of equations depending on the parameter~$ L $, we consider a subset for which the condition~(\ref {b=0}) is satisfied:
\begin{equation}\label{sin=0}
L=\frac{\pi n c}{2\omega_{0}\alpha}.
\end{equation} 
In this case, the system of equations~~\eqref{char1} is greatly simplified:
\begin{equation}\label{a-b-2}
\left\{\begin{array}{r} 
{\alpha ^{4} -6\alpha ^{2} \gamma ^{2} +\gamma ^{4} -2 \left(\alpha ^{2} -\gamma ^{2} \right)+ 
	\left[1-e^{2\omega_0 \gamma \tau } (-1)^n \right]=0;}\\ 
{4\alpha \gamma \left(\alpha ^{2} -\gamma ^{2} - 1 \right)  =0.} 
\end{array}\right.
\end{equation}

For $ \gamma \not = 0 $ we have
\begin{equation}\label{b-nonzero}
\gamma =\eta \sqrt{\alpha ^{2} - 1} , \qquad \quad   \left( \eta=\pm 1, \quad \alpha > 1\right).
\end{equation}
where $\eta=\pm 1$.

Substituting this expression for~$ \gamma $ into the first of the equations~(\ref {a-b-2}), we find
\begin{equation}\label{a-2}
\left(-1\right)^{n} \, e^{2\omega_0\gamma \tau } =4\alpha^2 \left(1 - \alpha^2 \right). 
\end{equation}

\begin{figure}[b]
	\includegraphics[width=1.2\linewidth]{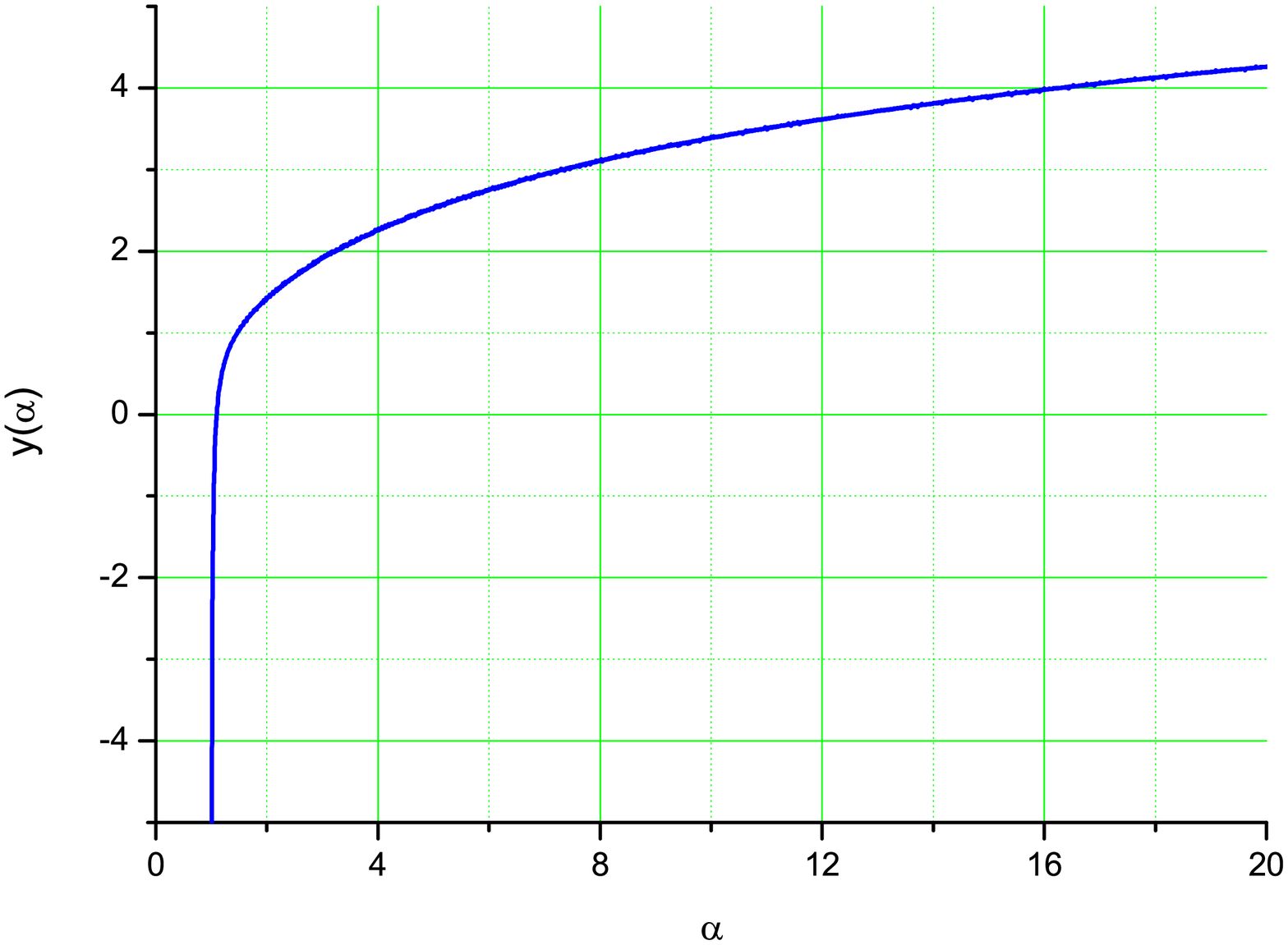}
	\caption{The plot of the function $y(\alpha)=\frac{1}{\pi}\ \frac{\alpha\,\ln \left[4\left( \alpha^4 - \alpha^2\right)  \right]}{\sqrt{\alpha^2-1}}$.}
	\label{fig:oscillator}
\end{figure}

Since $ \alpha > 1 $, $ n $ in this equation is an odd number ($ n = 2s + 1 $) and the equation is reduced to
\begin{equation}\label{ln}
(2s + 1) \eta = \frac {1} {\pi} \; \frac {\alpha \, \ln \left[4 \left(\alpha^ {4} - \alpha^{2} \right) \right]} {\sqrt {\alpha^{2} -1}}.
\end{equation}
The right side of this equation is a monotonically increasing function into the interval $ (1, \infty) $. The range of values of this function fills the entire interval $ (- \infty, \infty) $. Therefore, for each value of $ \eta \left(2s + 1 \right) $, the equation (\eqref {ln}) has a unique solution~$ \alpha_s (\eta) $.
The graph of the function contained in the right-hand side of this equation is shown in~Fig.\ref{fig:oscillator}.

\section{Difference between past and future}

The roots of the characteristic equation~\eqref{char2}, represented in Fig.\ref{fig:oscillator}, form an infinite set of complex numbers of the form\begin{equation}\label{Lambda}
\Lambda_{s}\left(\eta \right) =   \alpha_{s}\left( \eta\right) +i \eta \sqrt{\alpha_{s}^{2}\left( \eta\right) -1}  
\end{equation}
Each of the roots corresponds to a particular solution of the system of equations~\eqref{Eq-2}
\begin{equation}\label{part-sol}
x_{k,s}\left( t\right) = C_{k,s}\, e^{i\omega_{0}\alpha_{s}\left( \eta\right)\, t}\, e^{-\omega_{0}\eta \sqrt{\alpha_{s}^{2}\left( \eta\right) -1}\,t}.
\end{equation}
The amplitude of non-stationary oscillations corresponding to this particular solution is determined by the factor
\begin{equation}\label{amplitude}
A_{s}\left(\eta,\, t \right) = C_{k,s}\, e^{-\omega_{0}\eta \sqrt{\alpha_{s}^{2}\left( \eta\right) -1}\,t} 
\end{equation} 
and essentially depends on the parameter $ \eta $, which takes the values $ \pm 1 $.

\subsection{Future: Amplitudes of non-stationary oscillations at $ t > 0 $}

Let us consider in detail both cases: $ \eta = + 1 $ and $ \eta = -1 $.
\begin{enumerate}
	\item $ \eta = +1 $. In this case, the sequence $ \alpha_ {s} (1) $ increases monotonically ($ \alpha_ {s + 1} (1)> \alpha_ {s} (1) $), all the terms  of this sequence are noticeably more than one and $ \lim \limits_ {s \to \infty} \alpha_ {s} (1) = \infty $. Therefore, for $ t \to + \infty $, all amplitudes $ A_{s} \left(1, t \right) $ tend to zero.
	\item $ \eta = -1 $. In this case, all $ \alpha_ {s} (- 1) $ are practically equal to unity and the amplitudes of the corresponding oscillations increase slowly. It is essential that due to the increase in the amplitudes of oscillations with time, the condition of small oscillations~\eqref {small} begins to violate, the value of the interaction retardation~$ \tau $ ceases to be a constant and the problem becomes much more complicated.
\end{enumerate}

\subsection{Past: Amplitudes of non-stationary oscillations at $ t < 0 $}

The time reversal operation $ t \to -t $ does not affect the type of particular solutions like~\eqref{part-sol}, so to study the evolution of the oscillator, one must replace $ t \ to -t $ in these particular solutions.

Again, consider the cases $ \eta = 1 $ and $ \eta = -1 $, setting $ t <0 $.
\begin {enumerate}
\item $ \eta = +1 $. As before, the sequence $ \alpha_ {s} (1) $ increases monotonically ($ \alpha_ {s + 1} (1)> \alpha_ {s} (1) $>1), all terms of this sequence are more than one and $ \lim\limits_ {s \to \infty} \alpha_ {s} (1) = \infty $. Therefore, for $ t \to - \infty $, all the amplitudes $ A_{s} \left(1, t \right) $ tend to infinity.
\item $ \eta = -1 $. In this case, all $ \alpha_ {s} (- 1) $ are practically equal to unity and  the amplitudes of the corresponding oscillations slowly decrease.
\end {enumerate}

\section{Discussion}

The delay in the interaction between the constituents of a two-body oscillator leads to the following effects:
\begin{enumerate}
	\item Stationary oscillations exist only for a discrete set of equilibrium distances between the particles, determined by the condition~(\ref{b=0}). The frequencies of all stationary oscillations~$\omega$ are the same:
	\begin{equation}\label{freq-stat}
	\omega = \sqrt{2}\, \omega_0.
	\end{equation}	
	Condition~(\ref{b=0}) can be represented in an equivalent form
	\begin{equation}\label{b=0-2}
	L = \frac{n}{4} \lambda,\quad \lambda = \frac{2\pi c}{\omega},
	\end{equation} 
	where $ \lambda $ is the wavelength corresponding to the oscillation frequency~(\ref{freq-stat}).   
	\item If the equilibrium distance~$L$ between the particles does not satisfy the condition~(\ref{b=0}), then there are only non-stationary oscillations in the system.
	Depending on the sign of $\gamma = \mathrm{Im}\, \Lambda $, the corresponding oscillations either decay or increase. In both of these cases, the energies of the oscillations are not conserved.
	For~$\gamma > 0$, the corresponding oscillations are damped and the mechanical energy of these oscillations decreases. For~$\gamma <0$, the amplitude of the corresponding oscillations increases exponentially and the assump\-tions~(\ref{small}), (\ref{tau}) are no longer applicable. In particular, $\tau$ is no longer a constant~(\ref{tau}), but a function that depends on the solution of the system of equations~(\ref{Eq-2}). Thus, the delay of interactions between the particles performs functions similar to those of a thermodynamic reservoir, which, however, is not something external, but is inextricably linked to the particles.
	\item Especially it should be noted that even under the condition~(\ref{b=0}), in addition to stationary oscillations, the non-stationary oscillations with different frequencies corresponding to the complex roots of the characteristic equation may also exist~(\ref{char2}):
	\begin{equation}\label{char3}
	\Lambda_{s}\left(\eta \right) =   \alpha_{s}\left( \eta\right) +i \eta \sqrt{\alpha_{s}^{2}\left( \eta\right) -1} 
	\end{equation} 
	\item 
	Thus, a two-particle oscillator with delayed interactions has an infinite spectrum of oscillations, including both stationary and non-stationary oscillations. This spectrum depends essentially on the equilibrium distance~$L$ between the particles. 
\end{enumerate}

Thus, taking into account the seemingly insignificant but always existing delay in the interaction between the particles leads to a qualitative change in the dynamics of the system of interacting particles, including the phenomenon of irreversibility. Therefore, the classical theory of systems of particles with retarded interactions can be used for a correct microscopic justification of thermodynamics. Such an approach has certain advantages over the probabilistic foundation of classical thermodynamics of Boltzmann and Gibbs.

We omit the enormous difficulties in the practical use of the Gibbs method and the other approaches connected with it: these are mainly mathematical difficulties, such as the calculation of partition functions (both with realistic and with the simplest model interatomic potentials), the uncoupling of the BBGKY hierarchy, estimations of accuracies in approximate calculations, and so on. We will discuss only the most fundamental problems.

The microscopic foundation of thermodynamics within the framework of Gibbs statistical mechanics without establishing of the source of system stochastization is not satisfactory because of the numerous paradoxes generated by the use of probabilistic concepts in combination with deterministic classical mechanics. Numerous attempts to explain the irreversibility on the basis of the ergodic theory linking probability and thermodynamics have never led to a clear understanding of the nature of irreversibility, at least within the framework of classical mechanics.

We note some attempts to search for the mechanisms of the irreversible behavior of mechanical systems beyond the theory of probability.
\begin{enumerate}
	\item In 1940, a very interesting work was published by Synge~\cite{Synge}. In this paper Synge solved ``the electromagnetic two-body problem, based on the hypotheses (i) that the bodies are particles, (ii) that the fields are given by the retarded potential, (iii) that the force acting on a particle is the Lorentz ponderomotive force without a radiation term. It is found that energy disappears from the motion, so that the orbital particle slowly spirals in''. In fact, in this paper, the irreversible behavior of a two-body system with delayed interaction is proved.
	\item The idea of the non-uniqueness of the Cauchy problem for many-body systems as the cause of irreversibility was proposed in the paper~\cite{Kosyakov}. Hence it was concluded that the non-uniqueness of the solution of the Cauchy problem leads to an indeterministic behavior of the system. As it can be seen from our solution of the two-body oscillator model with delayed interaction, the deterministic and at the same time irreversible behavior of the system is provided by a more complete setting of the initial conditions: it is necessary to take into account the state of the system not only at the initial moment of time, but also its states at earlier time intervals.
	\item It should be pointed out on the recent papers~\cite{Lucia1,Lucia2},  in which the mechanism of macroscopic irreversibility as a consequence of microscopic irreversibility was developed. In these papers, atoms and molecules are considered as open systems being in continuous interaction with photons flowings from their surroundings. 
\end{enumerate}

\section{Conclusions}

The exact solution of a simple oscillator model with delayed interactions between the constituents leads to the following conclusions.
\begin{enumerate}
	\item The irreversible behavior of systems of interacting particles is a common property for both few-body and many-body classical system, having its origin  in the delay of interactions. 
	\item The unavoidable delay of interactions is sufficient for the irreversible behavior of the systems. The systems are irreversible in itself –-  there is no need to use any probabilistic hypotheses or other assumptions to explain the phenomenon of irreversibility in the systems of interacting particles. 
\end{enumerate}

	I am grateful to Profs. Ya. I. Granovsky and V. V. Uchaikin for useful discussions.

\end{document}